\documentclass[11pt,twoside]{article}

\usepackage{asp2004}
\usepackage{epsf}
\usepackage{psfig}
\usepackage{lscape}

\begin{document}
\title{Evolution of the 3.6$\micron$ Cluster Luminosity Function}   
\author{Adam Muzzin}   
\affil{University of Toronto/Spitzer Science Center}    
\author{Gillian Wilson,  Mark Lacy}   
\affil{Spitzer Science Center}    
\begin{abstract} 
We have measured the 3.6$\micron$ cluster luminosity function (LF) using a sample of 123 galaxy clusters selected from the 4 degree$^2$ {\it Spitzer} First Look Survey (FLS).  The clusters were selected on the basis of their R - 3.6$\micron$ colors using the cluster red-sequence technique of Gladders \& Yee (2000).  The binned LFs are well-fit by a Schechter function at all redshifts. However, we note two interesting trends.  Firstly, the evolution of M$_{*}$ with redshift is consistent with models that form galaxies in a single-burst of star-formation at {\it z} $>$ 2.0, and evolve passively thereafter.  Secondly, the faint-end slope of the LF appears to become shallower at higher redshift.  
We conclude that the most massive galaxies were formed in the cluster at high redshift ({\it z} $>$ 2.), while lower-mass galaxies have subsequently been accreted from the field at lower redshift ({\it z} $\sim$ 0.5).  These results are consistent with the ``downsizing'' picture seen in previous cluster studies using smaller samples. 
\end{abstract}
\section{Introduction}
The near-infrared cluster luminosity function is an important
diagnostic tool for understanding the evolution of cluster galaxies.  Using ground-based K-band data and a
sample of 38 clusters, de Propris et al., (1999) showed that the evolution of M$_{*}$ is consistent with a passively evolving population which formed at high redshift.  From a deep observation of one {\it z} = 1.27 cluster, Toft et al. (2004) showed that the faint end slope of the LF was shallower at higher redshifts. 
Our 3.6$\micron$ LFs confirm both of these results.
\section{Luminosity Functions}
We constructed LFs for our sample by stacking clusters in redshift intervals of $\delta z = 0.1$.
We measure number counts from the full 4 degree$^2$ of the FLS and use them to statistically subtract the background.
In order to stack clusters we artificially redshift each cluster to the median redshift of the bin.  
This involves computing the offset distance modulus as well as a small K-correction.  We use all galaxies within a radius of 750 Mpc to construct the LFs, and apply a small completeness correction calculated from Lacy et al. (2005).  The result is 10 LFs which span the redshift range 0.15 $< z <$ 1.22, and contain 7 - 21 clusters per bin.  Figure 1 shows the {\it z} = 0.15, 0.56, 0.83, \& 1.00 LFs.

\begin{figure}[!ht]
\begin{center}
\scalebox{0.65}{
 \includegraphics[25,350][575,730]{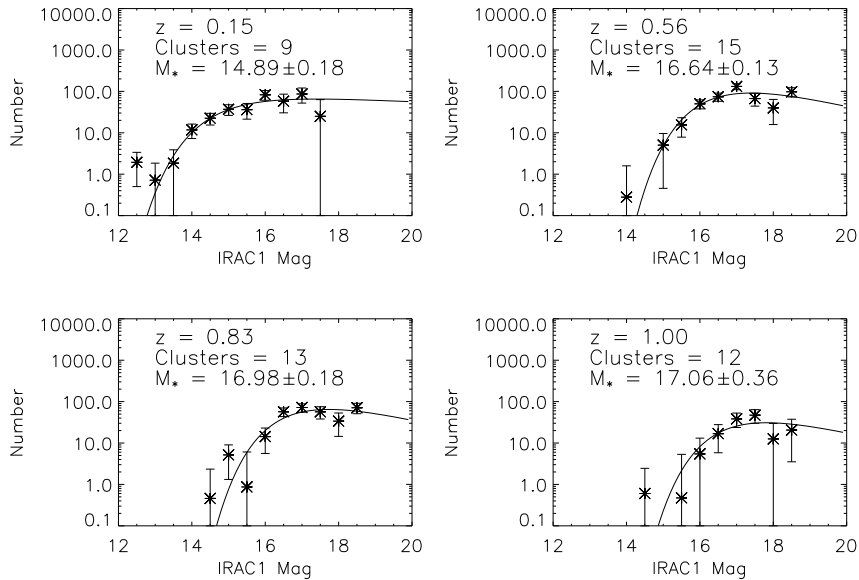}		
}
\caption{\footnotesize Example of mean LFs for four of the ten redshift intervals.  The solid line is the best-fit Schechter function.  The value of M$_{\star}$ and the number of clusters averaged to create the LF is listed in each panel.
}
\end{center}
\end{figure}

The FLS data is fairly shallow ($\sim$ M$_{*}$ + 1.5 at {\it z} =
1.0), and it does not allow a robust fit of the faint-end slope
($\alpha$) of the LFs.  Therefore, we chose to hold $\alpha$ constant
(adopting $\alpha = -0.9$, the value assumed by de Propris et
al. 1999). We fit only the normalization ($\phi_{*}$) and
characteristic magnitude (M$_{*}$).  At {\it z} $<$ 0.5, the LFs are
well fit (reduced $\chi^2$ $\sim$ 1). However at {\it z} $>$ 0.5 the
fits become much poorer.  In an attempt to improve the fit we allowed
$\alpha$ to vary in value from -0.2 to -1.5, in increments of 0.1.  We
found that $\alpha$ = -0.5 at $z > 0.5$, provided the best fit (mean
reduced $\chi^2$ of 0.83 compared to 1.1 for $\alpha$ = -0.9).  This
faint-end slope is consistent with the value of $\alpha$ = -0.64 found by Toft et al. (2004), and indicates that the number of low-mass galaxies in clusters decreases with redshift.  Furthermore, our fitted values of M$_{*}$ are consistent with models of passively evolving galaxies that form the bulk of their stars in a single-burst at {\it z} $>$ 2.  Our interpretation of these results is that the most massive galaxies in clusters form synchronously with the cluster at {\it z} $>$ 2, whereas many of the low-mass systems are accreted from the field at lower redshift ({\it z} $<$ 0.5).


\end{document}